\documentclass[useAMS]{mn2e}
\usepackage{graphicx}
\usepackage{psfig}
  
\def\kms{\relax \ifmmode {\,\rm km\,s}^{-1}\else \,km\,s$^{-1}$\fi}

\def\mincir{\ \raise-2.truept\hbox{\rlap{\hbox{$\sim$}}\raise5.truept
    \hbox{$<$}\ }}
\def\magcir{\ \raise-2.truept\hbox{\rlap{\hbox{$\sim$}}\raise5.truept
    \hbox{$>$}\ }}

\def\arcsec{\hbox{$^{\prime\prime}$}}

\def\sii{[S {\sc ii}]}
\def\heii{He{\sc ii}}

\def\oiii{[O {\sc iii}]}
\def\ha{H$\alpha$}
\def\hb{H$\beta$}

\def\te{$T_e$}

\begin{document}
\title[Discovery of the farthest known symbiotic star]{Discovery in IC10 of the farthest known symbiotic star
\thanks{Based on
    observations obtained at the Gemini Observatory, which is operated by
    the Association of Universities for Research in Astronomy, Inc., under a
    cooperative agreement with the NSF on behalf of the Gemini
    partnership. We also retrieved UBVRI images of the \lq\lq Survey of Local Group Galaxies 
    Currently Forming Stars", Massey et al. (2007).}}

\author[Gon\c calves, et al.]
  {Denise R. Gon\c calves$^{1}$\thanks{E-mail: denise@ov.ufrj.br}; 
  Laura Magrini$^{2}$; Ulisse Munari$^{3}$; Romano L. M. Corradi$^{4,5}$; 
\newauthor
  Roberto D. D. Costa$^{6}$ 
\\
  $^{1}$ UFRJ - Observat\'orio do Valongo, Ladeira Pedro Antonio 43, 20080-090 Rio de Janeiro, Brazil\\
  $^{2}$ INAF - Osservatorio Astrofisico di Arcetri, Largo E. Fermi 5, I-50125 Firenze, Italy\\
  $^{3}$ INAF - Osservatorio Astronomico di Padova, via dell'Osservatorio 8, 36012 Asiago (VI), Italy\\
  $^{4}$ Isaac Newton Group of Telescopes, Ap.\ de Correos 321, E-38700 Santa Cruz de la Palma, Spain\\
  $^{5}$ Instituto de Astrof\'\i sica de Canarias, E-38205 La Laguna, Tenerife, Spain\\ 
  $^{6}$ IAG - Universidade de S\~ao Paulo, Rua do Mat\~ao 1226, 05508-900 S\~ao Paulo, Brazil} 


\date{Accepted ?. Received ?; in original form ?}

\pagerange{\pageref{firstpage}--\pageref{lastpage}} \pubyear{2008}

\maketitle

\label{firstpage}

\begin{abstract}

We report the discovery of the first known symbiotic star in IC10, a
starburst galaxy belonging to the Local Group, at a distance of $\sim$750~kpc. 
The symbiotic star was
identified during a survey of emission-line objects. It shines at
$V$=24.62$\pm$0.04, $V$$-$$R_{\rm C}$=2.77$\pm$0.05 and $R_{\rm
C}$$-$$I_{\rm C}$=2.39$\pm$0.02 and suffers from $E_{B-V}$=0.85$\pm$0.05
reddening. The spectrum of the cool component well matches that of
solar neighborhood M8III giants. The observed emission lines belong to Balmer series,
[SII], [NII] and [OIII]. They suggest a low electronic density, negligible
optical depth effects and 35,000$<$$T_{\rm eff}$$<$90,000~K for the ionizing
source.  The spectrum of the new symbiotic star in IC10 is an almost
perfect copy of that of Hen~2-147, a well known Galactic symbiotic star and
Mira.

\end{abstract}


\section[]{Introduction}

The importance of symbiotic stars to understand stellar evolution in binary
systems and the origin of type Ia supernovae is widely recognized (eg.
Corradi, Miko\l ajewska \& Mahoney 2003). However, so far only a few symbiotic stars are known
in external galaxies, all belonging to the inner Local Group: eight in the
LMC, six in the SMC and one in Draco (Belczy\'nski et al. 2000). The vast majority
of symbiotic stars known in our Galaxy has been discovered during objective
prism surveys. The need of spectroscopy to identify them helps to explain the
paucity of discoveries outside our Galaxy. In this {\em Letter} we report 
our spectroscopic discovery of the first symbiotic star known in
IC10, a dwarf starburst galaxy belonging to the Local Group. The partnership
with IC10 makes this symbiotic star, named hereafter IC10 SySt-1, the most
distant one known at this time, at a distance $\sim$750~kpc (the most
recent estimate of the distance to IC10 is 690 to 790 kpc by Kniazev, Pustilnik \& 
Zucker 2008).

\begin{figure*}
\begin{center}
\vbox{\psfig{file=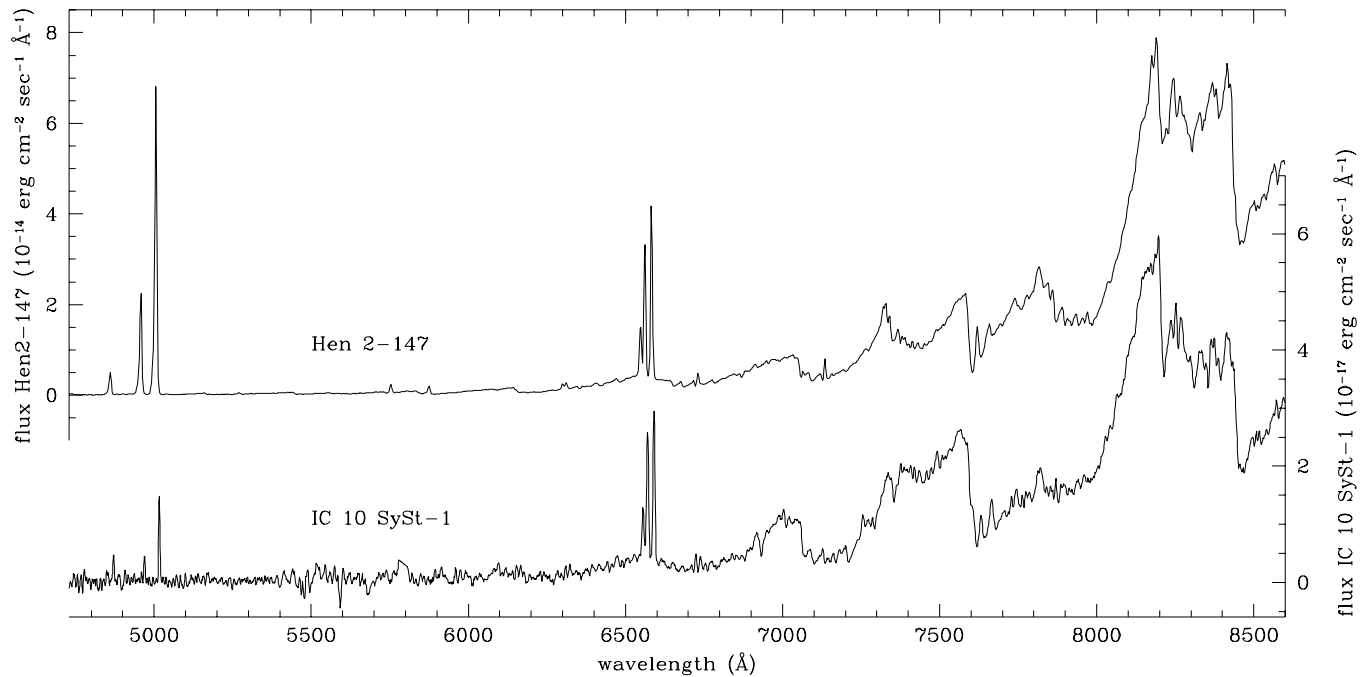,width=18.0truecm}}
\caption{The observed spectrum of IC10 StSy-1 as a composition of the blue
and red spectra obtained with GMOS-N. The spectra of the Galactic symbiotic 
star and Mira Hen~2-147 (from Munari and Zwitter 2002) is plotted for comparison.   
The spectrum of IC10 StSy-1 has been Gaussian filtered to reduce
the noise and match in resolution that of Hen~2-147.}
\end{center}
\end{figure*}

In a typical symbiotic star, matter is transferred from a cool giant to a
white dwarf companion, and stable H-burning at the surface of the latter
(Sokoloski 2003) provides the energy input to ionize some fraction of the
circumstellar gas. The stable H-burning processing of the accreted matter turns
symbiotic stars into a viable progenitor channel for the type Ia supernovae
(Munari and Renzini 1992). The importance of such a channel depends on
several factors, like the efficiency in mass growth of the white dwarf, the
mass reservoir in the donor star and the efficiency of the mass transfer,
the partnership with the old population, and the number of symbiotic stars
per unit mass. The total number of symbiotic stars in our Galaxy has been
variously estimated as 4$\times$10$^5$ by Magrini et al. (2003),
3$\times$10$^5$ by Munari and Renzini (1992), 3$\times$10$^4$ by Kenyon et
al. (1993) and 3$\times$10$^3$ by Allen (1984). The large differences among
them arise mainly in the way the incompleteness discovery fraction is computed 
starting from the small observed sample ($\sim$2$\times$10$^2$, Belczynski et al. 2000).
Surveys for and discovery of symbiotic stars in other galaxies should help
to better evaluate the incompleteness discovery fraction in our own Galaxy.

We performed a survey of emission-line populations in IC10, to
search for and to study the spectroscopic properties of planetary nebulae
and HII regions.  This was made with the Gemini North telescope, by
means of H$\alpha$ narrow-band continuum-subtracted imaging and followup
spectroscopy.  During this survey we discovered the new symbiotic star whose
basic properties are described in the following sections.

\section[]{Observational data}

\subsection[]{Imaging}

IC10 StSy-1 was identified as an H$\alpha$ emitter in images obtained with
the 8.1-m Gemini North Telescope and its Multi-Object Spectrograph (GMOS-N).
Images were taken in queue mode on August 08, 2007. The field of view was
5\arcmin$\times$5\arcmin, centered at $00^h20^m23\fs16$
$+59\degr17\arcmin34\farcs7$.

A continuum-subtracted H$\alpha$ image was built using two narrow-band
frames, one taken with the H$\alpha$ filter Ha$_-$G0310 centered at 6550 \AA\ and
70~\AA\ wide, and the other with the H$\alpha$-continuum filter HaC$_-$0311,
centered at 6620~\AA\ and 70~\AA\ wide. The exposure times were 400~s, split in
two sub-exposures, through both filters.

\subsection[]{Spectroscopy}

Spectra of IC10 StSy-1 were obtained in queue mode with GMOS-N at Gemini
North, using two different gratings: R400+G5305 (`red'), with 3 exposures of
1,700s each, on October 11, 2007, and B600+G5303 (`blue') with
4$\times$1,700s exposures, on October 14 and 18, 2007. The slit width was
1~arcsec, and the pixel binning were 2$\times$2 (spectral$\times$spatial). 
The spatial scale and reciprocal dispersions of the spectra were as follows:
0\farcs094 and 0.3~nm per binned pixel, in `blue'; and 0\farcs134 and
0.8~nm per binned pixel, in `red'. Seeing varied from
$\sim$0.5\arcsec\ to
$\sim$0.6\arcsec\ for the R400 spectra, and it was $\sim$0.6\arcsec\ for the
two runs in which B600 spectra were taken. CuAr lamp exposures were obtained
with both gratings for wavelength calibration.  The effective `blue' plus
`red' spectral coverage of the IC10 StSy-1 spectrum was from 3700~\AA\ to
9500~\AA. 

The data were reduced and calibrated using the  Gemini {\sc gmos data
reduction script} and {\sc long-slit} tasks, both being part of IRAF. 
Spectra of the spectrophotometric standard Wolf1346 (Massey et al. 1988,
Massey and Gronwall 1990), obtained with the same instrumental setups as 
IC10~SySt-1 on two different nights (September 17 and October 5, 2007), 
were used to calibrate our spectra of
IC10~SySt-1. This allowed to recover the actual slope of the spectrum,
although not its flux zero point. The latter was obtained by imposing that
the $R_{\rm C}$-band flux integrated over the spectrum would match the
$R_{\rm C}$=21.85 mag we measured from direct VR$_{\rm C}$I$_{\rm C}$
photometry of IC10~SySt-1 (see next sect.~3.1). The spectrum of IC10~SySt-1
so calibrated is presented in Figure~1.

\section{Discussion}

\subsection[]{Astrometric position and VR$_{\rm C}$I$_{\rm C}$ magnitudes}

Massey et al. (2007) obtained, with the Kitt Peak National Observatory
and Cerro Tololo Inter-American Observatory 4-m telescopes and Mosaic
cameras, UBVR$_{\rm C}$I$_{\rm C}$ photometry of the resolved stellar
population of several dwarf galaxies with active star formation, including
IC10. Massey et al. (2007) did not provide photometry for IC10 StSy-1. We have 
used their source observations (accessible via
ftp\footnote{ftp://ftp.lowell.edu/pub/massey/lgsurvey}), and derived for
IC10 StSy-1 $V$=24.62$\pm$0.04, $V$$-$$R_{\rm C}$=2.77$\pm$0.05 and $R_{\rm
C}$$-$$I_{\rm C}$=2.39$\pm$0.02 via aperture photometry. The symbiotic star
is below detection threshold both in $U$- and $B$-band frames.

\begin{figure}
\begin{center}
\vbox{\psfig{file=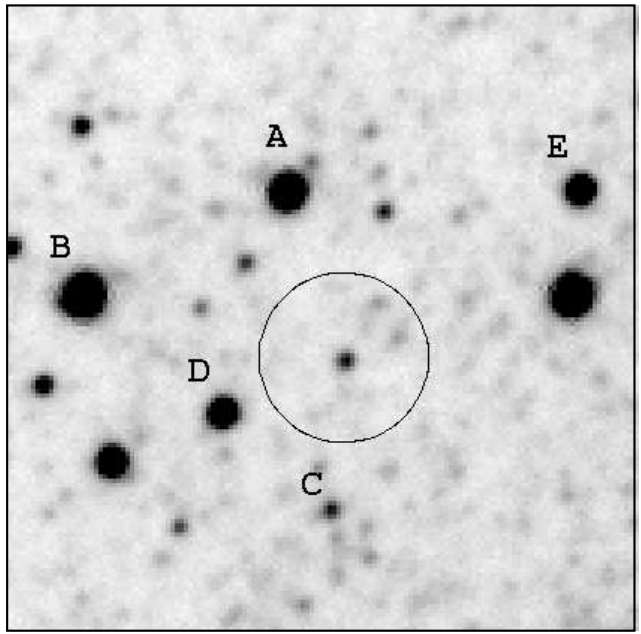,width=8.5truecm}}
\caption{$I_{\rm C}$-band finding chart covering a field of view of
37$\times$37 arcsec from Massey et al. (2007) observations. North is up and 
East is to the left. IC10 StSy-1 ($00^h20^m33\fs59$ $+59\degr18\arcmin45\farcs9$) is marked with a 
circle. The photometry of the stars marked with A to E is given in  Table~1.}
\end{center}
\end{figure}

The position we derived for IC10~SySt-1 on Massey et al. (2007)
astrometrically calibrated frames is RA=00:20:33.59 and DEC=+59:18:45.9
(J2000.0), with an error smaller than 1 arcsec.
This position is marked on the $I_{\rm C}$-band finding chart presented
in Figure~2.  It covers a field of view of 37$\times$37 arcsec and
shows stars down to $I_{\rm C}$=20~mag. We also marked on Figure~2 a
few field stars suitable to serve as local photometric standards, 
whose photometry is given in Table~1.

\begin{table}
\caption{Magnitudes of the finding chart stars of Figure~2.}
\centering
\begin{tabular}{lllll}
\hline
\multicolumn{5}{c}{}\\
  &  V               & B-V             & V-R$_C$         & R-I$_C$   \\
\multicolumn{5}{c}{}\\
A & 17.509$\pm$0.005 & 0.992$\pm$0.005 & 0.587$\pm$0.005 & ...\\
B & 16.983$\pm$0.005 & 1.200$\pm$0.005 & 0.688$\pm$0.005 & ...\\
C & 22.185$\pm$0.012 & 1.920$\pm$0.020 & 1.243$\pm$0.014 & 1.425$\pm$0.008 \\
D & 18.909$\pm$0.005 & 1.274$\pm$0.005 & 0.794$\pm$0.005 & 0.822$\pm$0.005 \\
E & 18.575$\pm$0.005 & 1.110$\pm$0.005 & 0.652$\pm$0.005 & 0.606$\pm$0.005 \\
\multicolumn{5}{c}{}\\
\hline
\end{tabular}
\end{table}

\subsection[]{Classification}

The spectrum of IC10~SySt-1 presented in Figure~1 is that of a symbiotic star:
strong molecular TiO absorption bands from a cool giant with superimposed
emission lines tracing the presence of a hot companion that ionizes the
circumstellar nebula fed by the mass loss from the giant. The presence of
[OIII] requires a temperature of the photo-ionization source in excess of
35,000~K, which satisfy the classification criteria for symbiotic stars
adopted by Belczy\'nski et al. (2000). Our spectrum is too underexposed at
the shortest wavelengths to distinguish between absence or presence of
HeII~4686 emission line, which would be required for the more stringent
classification criteria among symbiotic stars adopted by Allen (1984) and
that would trace ionization temperatures in excess of 55,000~K.

The partnership of IC10~SySt-1 with symbiotic stars is reinforced by the
comparison, carried out in Figure~1, with He~2-147. This is a bona fide
symbiotic star containing a long period Mira and a toroidal expanding
circumstellar nebula resolved in both ground-based and HST observations
(Munari and Patat 1993, Corradi et al. 1999, Santander-Garc\'\i a et al.
2007). The only significant difference with Hen~2-147 in Figure~1 is a lower
intensity in IC10~SySt-1 of [OIII] lines with respect to [NII], probably
tracing a lower temperature of the photo-ionizing source. A list and
integrated fluxes of the emission lines identified in the spectrum of
IC10~SySt-1 is given in Table~2.

\begin{table}
\caption{Emission lines observed in IC10 StSy-1. $F$: observed fluxes
in units of 10$^{-17}$ erg cm$^{-2}$ sec$^{-1}$ (errors $\pm$10\%).  
$F_\circ$: reddening corrected fluxes scaled to H$\beta$}
\centering
\begin{tabular}{llrr}
\hline
\multicolumn{4}{c}{}\\
&&  $F$  & $F_\circ$   \\
\multicolumn{4}{c}{}\\
 4861    &\hb    &  2.2  &1.00 \\
 4959    &[OIII] &  2.2  &0.91 \\
 5007    &[OIII] &  7.1  &3.08 \\
 6548    &[NII]  &  7.0  &1.13 \\
 6563    &\ha    & 18.7  &2.85 \\
 6584    &[NII]  & 21.0  &3.25 \\
 6717    &]SII]  &  2.1  &0.28 \\
 6731    &[SII]  &  1.9  &0.24 \\
\multicolumn{4}{c}{}\\
\hline
\end{tabular}
\end{table}
 
\subsection[]{Spectral type and reddening}

\begin{figure*}
\begin{center}
\vbox{\psfig{file=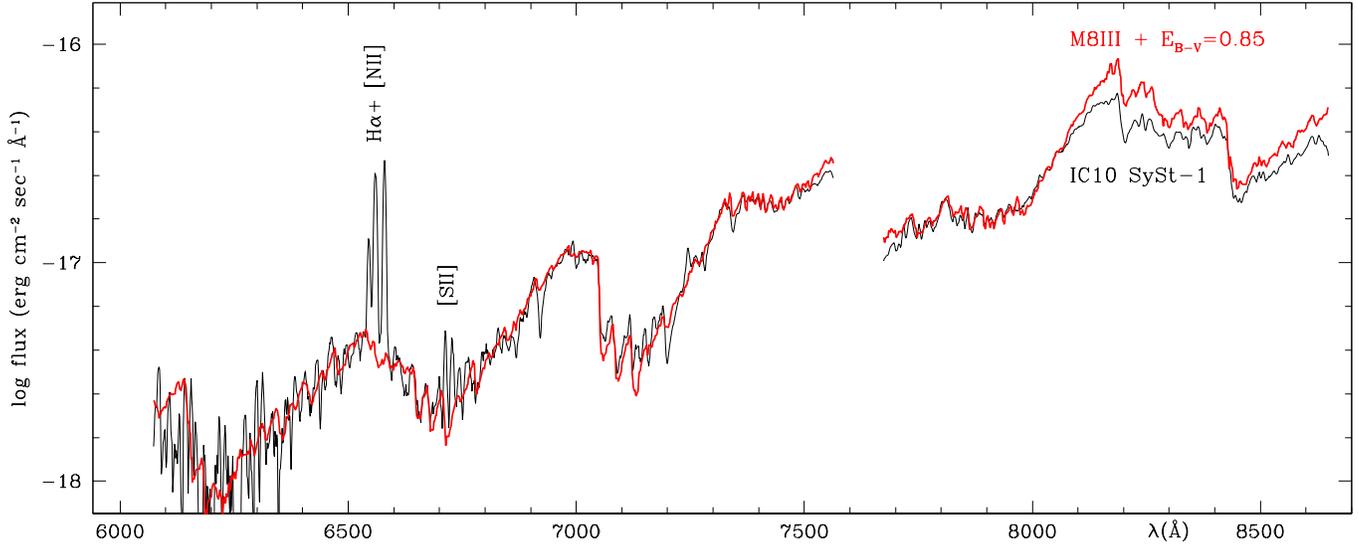,width=18.0truecm}}
\caption{Comparison of an M8III spectrum from Fluks et al. (1994, thicker line),
reddened by $E_{B-V}$=0.85 with our spectrum of IC10 SySt-1 from Figure~1 
(thinner line).}
\end{center}
\end{figure*}

To derive the spectral type of the cool giant, we have compared the red
portion of the IC10~SySt-1 spectrum with the spectral atlas of Fluks et al.
(1994), which is based on observation of solar neighborhood objects. We obtained 
an excellent match with a M8III type, and the much poorer respective fits
exclude a classification as either M7 or M9 spectral types. The comparison
with the unreddened spectra of Fluks et al. (1994) also firmly constrains the
reddening, found to be $E_{B-V}$=0.85$\pm$0.05. The fit to the observed
spectrum of IC10~SySt-1 with an M8III from Fluks et al. (1994) and reddened
by $E_{B-V}$=0.85 is presented in Figure~3. Integrating the BVR$_{\rm
C}$I$_{\rm C}$ pass-bands would provide B=26.79, V=24.43, R$_{\rm C}$=21.85
and I$_{\rm C}$=19.33 for the reddened M8III fitting spectrum, and B=23.47,
V=21.95, R$_{\rm C}$=20.12 and I$_{\rm C}$=17.94 for the unreddened one.

The reddening we found for IC10~SySt-1 is the same as reported by Mateo (1998)
for the IC10 as a whole, suggesting a negligible circumstellar
contribution. The $E_{B-V}$=0.85 value for the reddening affecting
IC10~SySt-1 is also supported by the flux ratio of Balmer H$\alpha$
and H$\beta$ emission lines. Their recombination line ratio under Case B
conditions and negligible self-absorption is $\sim$2.9 (Osterbrock and Ferland 2006, 
their Table 4.4). 
In IC10~SySt-1 the observed ratio is 8.5 which reduces to 2.9 once corrected
for $E_{B-V}$=0.85.

\subsection[]{Electron density, oxygen abundance and temperature of the 
ionizing source}

A preliminary estimate of the nebular conditions can be derived from the 
data at hand (see Table~2), using the {\sc nebular} IRAF package (Shaw \& Dufour 
1994).

The electron density in the circumstellar nebula can be estimated as
400$\pm$200 cm$^{-3}$ using the \sii$\lambda\lambda$ 6716, 6731 \AA\
doublet. An upper limit to the flux of \oiii\ 4363~\AA\ emission line (6\%
of \hb ) gives a lower limit to the electron temperature of $\sim$17,000~K
from the I($\lambda$4959+$\lambda$5007)/I($\lambda$4363) ratio. Using this
limit for \te\oiii\ and the O$^{++}$ abundance, the lower limit to the total oxygen
abundance (Kingsburgh \&  Barlow 1994) 	
is 12 + $\log$(O/H) $>$ 7.4, the solar abundance being 8.66 (Asplund Grevesse \& Sauval 2005).   
An upper limit for the integrated flux of \heii\ 4686~\AA\ emission line  
amounting to 5\% of the integrated flux of \hb\ corresponds to an upper limit  
$T_{\rm eff}$$<$90,000~K for the temperature of the ionizing source,
 the lower limit $>$35,000~K being set by the presence of \oiii\ 
 (Kaler \& Jacoby 1989). 

Contrary to many symbiotic stars, the nebular lines of IC10 StSy-1 do
not indicate the presence of high density ionized gas. The observed
lines could instead originate in an extended,  low density nebula.  
To the low density conditions characterizing IC10 SySt-1 spectrum could be 
contributing the emission from external and extended nebular regions 
similar to those resolved around many symbiotic Miras in our Galaxy (Corradi 
et al. 1999), like the one around Hen~2-147, or the more spectacular ones 
of He2-104 (Corradi et al. 2001) and R Aqr (Gon\c calves et al. 2003).
\section*{Acknowledgments}

We would like to thank Peter McGregor, the referee, for his suggestions.  
Two Brazilian and one Italian agency gave us partial support for this work. 
So DRG and LM would like to thank FAPESP (2003/09692-0 and 2006/59301-6, respectively) 
and the EC Research Training Network MRTN-CT-2006-035890 ``Constellation''. 
DRG also thanks FAPERJ's (E-26/110.107/2008) grant.

\bsp

\label{lastpage}

\end{document}